\documentclass[12pt,english]{article}
\pdfoutput=1
\usepackage{graphicx,array}
\usepackage{cite}
\usepackage{color}
\usepackage{latexsym}
\usepackage{amsthm}
\usepackage{amsmath}
\usepackage[titletoc]{appendix}
\usepackage{enumitem}
\usepackage{amssymb}
\usepackage{lipsum}
\usepackage[unicode=true,
 bookmarks=true,bookmarksnumbered=false,bookmarksopen=false,
 breaklinks=false,pdfborder={0 0 1},backref=false,colorlinks=true]
 {hyperref}
\hypersetup{pdftitle={Mad-Dog Everettianism: Quantum Mechanics at Its Most Minimal},
 pdfauthor={Sean M. Carroll and Ashmeet Singh},
 citecolor=blue,linkcolor=blue,urlcolor=blue,citecolor=blue,linkcolor=blue}
\usepackage{breakurl}
\usepackage[hang,flushmargin]{footmisc} 
\usepackage{footnote}
\usepackage{footmisc}

\newcommand{\astfootnote}[1]{
\let\oldthefootnote=\thefootnote
\setcounter{footnote}{1}
\renewcommand{\thefootnote}{\fnsymbol{footnote}}
\footnote{#1}
\let\thefootnote=\oldthefootnote
}

\def\simgt{\mathrel{\lower2.5pt\vbox{\lineskip=0pt\baselineskip=0pt
           \hbox{$>$}\hbox{$\sim$}}}}
\def\simlt{\mathrel{\lower2.5pt\vbox{\lineskip=0pt\baselineskip=0pt
           \hbox{$<$}\hbox{$\sim$}}}}

\newcommand{\be}{\begin{equation}}
\newcommand{\ee}{\end{equation}}
\newcommand{\bea}{\begin{eqnarray}}
\newcommand{\eea}{\end{eqnarray}}

\newcommand{\Tr}{\operatorname{Tr}}

\newcommand{\HH}{\mathcal{H}}

\DeclareMathOperator{\co}{:}
\newcommand{\R}{\mathcal{R}}

\definecolor{nicered}{rgb}{0.7,0.1,0.1}
\definecolor{nicegreen}{rgb}{0.1,0.5,0.1}
\hypersetup{colorlinks,citecolor=black,linkcolor=black,urlcolor=black}

\newcommand\blfootnote[1]{%
  \begingroup
  \renewcommand\thefootnote{}\footnote{#1}%
  \addtocounter{footnote}{-1}%
  \endgroup
}

\begin{document}
\baselineskip=14pt
\hfill CALT-TH-2018-006
\hfill

\setcounter{footnote}{0}

\vspace{2cm}
\thispagestyle{empty}
\begin{center}
{\LARGE\bf
Mad-Dog Everettianism:\\ \vskip.2cm
Quantum Mechanics at Its Most Minimal}\\
\bigskip\vspace{1cm}{
{\large Sean M.\ Carroll and Ashmeet Singh$^{\dag}$\blfootnote{$^{\dag}$e-mail: \url{seancarroll@gmail.com, ashmeet@caltech.edu} \\ Submitted to the Foundational Questions Institute Essay Competition, ``What Is Fundamental?" \\Corresponding author: SMC.}}
} \\[7mm]
 {\it Walter Burke Institute for Theoretical Physics\\
    California Institute of Technology \\
    1200 E. California Blvd., Pasadena, CA 91125} \\
 \end{center}
\bigskip
\centerline{\large\bf Abstract}

\begin{quote} \small
To the best of our current understanding, quantum mechanics is part of the most fundamental picture of the universe.
It is natural to ask how pure and minimal this fundamental quantum description can be.
The simplest quantum ontology is that of the Everett or Many-Worlds interpretation, based on a vector in Hilbert space and a Hamiltonian.
Typically one also relies on some classical structure, such as space and local configuration variables within it, which then gets promoted to an algebra of preferred observables.
We argue that even such an algebra is unnecessary, and the most basic description of the world is given by the spectrum of the Hamiltonian (a list of energy eigenvalues) and the components of some particular vector in Hilbert space.
Everything else -- including space and fields propagating on it -- is emergent from these minimal elements.
\end{quote}

\vfill\eject

\begin{itemize}
\item \textbf{Taking quantum mechanics seriously}
\end{itemize}

The advent of modern quantum mechanics marked a profound shift in how we view the fundamental laws of nature: it was not just a new theory, but a new \emph{kind} of theory, a dramatic shift from the prevailing Newtonian paradigm.
Over nine decades later, physicists have been extremely successful at applying the quantum rules to make predictions about what happens in experiments, but much less successful at deciding what quantum mechanics actually \emph{is} -- its fundamental ontology and indeed its relation to underlying reality, if any.

One obstacle is that, notwithstanding the enormous empirical success of quantum theory, we human beings still tend to think in classical terms.
Quantum theory describes the evolution of a state vector in a complex Hilbert space, but we populate our theories with ideas like ``spacetime," ``particles," and ``fields."
We typically construct quantum theories by starting with some classical theory and then ``quantizing" it.
Presumably Nature works the other way around: it is quantum-mechanical from the start, and a classical limit emerges in the right circumstances.

In this essay we ask how far we can take the idea that the world is fundamentally quantum, with a minimal plausible ontology: a Hilbert space $\HH$, a vector $|\psi\rangle$ within it, and a Hamiltonian $\hat H$ governing the evolution of that vector over time.
This is a version of the Everettian (Many-Worlds) approach to quantum mechanics, in which the quantum state is the only variable and it smoothly evolves according to the Schr\"odinger equation with a given Hamiltonian,
\be
  \hat H |\psi(t)\rangle = i \partial_t|\psi(t)\rangle.
\ee
Our approach is distinguished by thinking of that state as a vector in Hilbert space, without any preferred algebra of observables, and without necessarily representing Hilbert space as a set of wave functions over some classical variables.
All of the additional elements familiar in physical theories, we will argue, can be emergent from the state vector (cf.\ \cite{Giddings:2015lla}).
We call this approach ``Mad-Dog Everettianism," to emphasize that it is as far as we can imagine taking the program of stripping down quantum mechanics to its most pure, minimal elements.\footnote{The name is inspired by philosopher Owen Flanagan's description of his colleague Alex Rosenberg's philosophy as ``Mad-Dog Naturalism."}

\begin{itemize}
\item \textbf{The role of classical variables}
\end{itemize}

The traditional way to construct a quantum theory is to posit some classical configuration space $\mathbf{X}$ (which could be momentum space or some other polarization of phase space), then considering the space of all complex-valued functions on that space.
With an appropriate inner product $\langle \cdot,\cdot\rangle$, the subset of square-integrable functions forms a Hilbert space:
\be
 \HH = \{\psi : \mathbf{X} \rightarrow \mathbb{C}\, {\big|} \langle\psi, \psi\rangle < \infty\}.
\ee
This gives us a \emph{representation} of $\HH$, but the Hilbert space itself is simply a complete, normed, complex-valued vector space. 
That gives us very little structure to work with: all Hilbert spaces of the same finite dimensionality are isomorphic, as are infinite-dimensional ones that are separable (possessing a countable dense subset, which implies a countable orthonormal basis).
We may therefore ask, once $\HH$ is constructed, is there any remnant of the original classical space $\mathbf{X}$ left in the theory?

The answer is ``not fundamentally, no."
A given representation might be useful for purposes of intuition or calculational convenience, but it is not necessary for the fundamental definition of the theory.
Representations are very far from unique, even if we limit our attention to representations corresponding to sensible physical theories.

One lesson of dualities in quantum field theories is that a single quantum theory can be thought of as describing completely different classical variables.
The fundamental nature of the ``stuff'' being described by a theory can change under such dualities, as in that between the sine-Gordon boson in 1+1 dimensions the theory of a massive Thirring fermion \cite{PhysRevD.11.2088}.
Even the dimensionality of space can change, as is well-appreciated in the context of the AdS/CFT correspondence, where a single quantum theory can be interpreted as either a conformal field theory in a fixed $d$-dimensional Minkowski background or a gravitational theory in a dynamical $(d+1)$-dimensional spacetime with asymptotically anti-de~Sitter boundary conditions \cite{Maldacena:1997re}.

The lesson we draw from this is that Nature at its most fundamental is simply described by a vector in Hilbert space.
Classical concepts must emerge from this structure in an appropriate limit.
The problem is that Hilbert space is relatively featureless; given that Hilbert spaces of fixed finite or countable dimension $D$ are all isomorphic, it is a challenge to see precisely how a rich classical world is supposed to emerge.

Ultimately, all we have to work with is the Hamiltonian and the specific vector describing the universe.
In the absence of any preferred basis, the Hamiltonian is fixed by its spectrum, the list of energy eigenvalues:
\be
  \{E_0, E_1, E_2, \ldots\}\ , \qquad \hat{H}|n\rangle = E_n|n\rangle\ ,
\ee
and the state is specified by its componenents in the energy eigenbasis,
\be
  \{\psi_0, \psi_1, \psi_2, \ldots\}\ ,\qquad |\psi\rangle = \sum_n \psi_n|n\rangle.
\ee
The question becomes, how do we go from such austere lists of numbers to the fullness of the world around us?

\begin{itemize}
\item \textbf{The role of emergence}
\end{itemize}

One might ask why, if the fundamental theory of everything is fixed by the spectrum of some Hamiltonian, we don't simply imagine writing the state of the universe in the energy eigenbasis, where its evolution is trivial?
The answer is the one that applies to any example of emergence: there might be other descriptions of the same situation that provide useful insight or computational simplification.

Consider the classical theory of $N$ particles moving under the influence of some multi-particle potential in $3$ dimensions of space.
The corresponding phase space is $6N$-dimensional, and we \emph{could} simply think of the theory as that of one point moving in a $6N$-dimensional structure.
But by thinking of it as $N$ particles moving in a $3$-dimensional space of allowed particle positions, we gain enormous intuition; for example, it could become clear that particles influence each other when they are nearby in space, which in turn suggests a natural way to coarse-grain the theory.
Similarly, writing an abstract vector in Hilbert space as a wave function over some classical variables can provide crucial insight into the most efficient and insightful way to think of what is happening to the system.

\begin{itemize}
\item \textbf{Local finite-dimensionality}
\end{itemize}

The Hilbert spaces considered by physicists are often infinite-dimensional, from a simple harmonic oscillator to quantum field theories. For separable Hilbert spaces (finite-dimensional or infinite-dimensional countable), the Stone-von~Neumann theorem guarantees us uniqueness of the irreducible representation of the algebra of the canonical commutation relations (CCRs), up to unitary equivalence. 
In non-separable Hilbert spaces, however, there can be unitarily inequivalent representations of the CCRs, implying that the physical subspaces spanned by eigenstates of operators in a particular representation will be different. 
In Algebraic Quantum Field Theory this is described by Haag's theorem \cite{haag55}. 
Then different choices of states (a unit-normed, positive linear functional) on the algebra specify different inequivalent (cyclic) representations. 
One might therefore conclude that specification of the algebra state is also an important ingredient in defining a quantum theory, over and above the Hamiltonian.

However, there are good reasons from quantum gravity to think that the true Hilbert space of the universe is ``locally finite-dimensional'' \cite{Bao:2017rnv}.
That is, we can decompose $\HH$ into a (possibly infinite) tensor product of finite-dimensional factors,
\be
  \HH = \bigotimes_\alpha \HH_\alpha,
  \label{decomp}
\ee
where for each $\alpha$ we have $\dim (\HH_\alpha) <\infty$.
If we have factored the Hilbert space into the smallest possible pieces, we will call these ``micro-factors."
The idea is that if we specify some region of space and ask how many states could possibly occupy the region inside, the answer is finite, since eventually the energy associated with would-be states becomes large enough to create a black hole the size of the region \cite{bekenstein1981}.
Similarly, our universe seems to be evolving toward a de~Sitter phase dominated by vacuum energy; a horizon-sized patch of such a spacetime is a maximum-entropy thermal state \cite{Carroll:2017kjo} with a finite entropy and a corresponding finite number of degrees of freedom \cite{Fischler2000,Banks2000}.

There are subtleties involved with trying to map collections of factors in (\ref{decomp}) directly to regions of space, including the fact that ``a region of space'' $\R$ might not be well-defined across different branches of the quantum-gravitational wave function.
All that matters for us, however, is the existence of a decomposition of this form, and the idea that everything happening in one particular region of space on a particular branch is described by a finite-dimensional factor of Hilbert space $\HH_\R$ that can be constructed as a finite tensor product of micro-factors $\HH_\alpha$.
Given some overall pure state $|\psi\rangle \in \HH$, physics within this region is described by the reduced density operator
\be
  \rho_\R = \Tr_{\bar\R} |\psi\rangle\langle\psi|.
\ee
In that case, there is no issue of specifying the correct algebra of observables: the algebra is simply ``all Hermitian operators acting on $\HH_\R$."
Any further structure must emerge from the spectrum of the Hamiltonian and the quantum state.

\begin{itemize}
\item \textbf{Spacetime from Hilbert space}
\end{itemize}

Fortunately, we are guided in our quest by the fact that we know a great deal about what an appropriate emergent description should look like -- a local effective field theory defined on a semiclassical four-dimensional dynamical spacetime.
The first step is to choose a decomposition of the Hilbert space $\HH_\R$ (representing, for example, the interior of our cosmic horizon) into finite-dimensional micro-factors.
We can say that the Hamiltonian is ``local'' with respect to such a decomposition if, for some small integer $k$, the Hamiltonian connects any specific factor $H_{\alpha_*}$ to no more than $k$ other factors; intuitively, this corresponds to the idea that degrees of freedom at one location only interact with other degrees of freedom nearby.

It turns out that a generic Hamiltonian will not be local with respect to \emph{any} decomposition, and for the special Hamiltonians that can be written in a local form, the decomposition in which that works is essentially unique \cite{Cotler:2017abq}.
In other words, for the right kind of Hamiltonian, there is a natural decomposition of Hilbert space in which physics looks local, which is fixed by the spectrum alone. 
From the empirical success of local quantum field theory, we will henceforth assume that the Hamiltonian of the world is of this type, at least for low-lying states near the vacuum.

This preferred local decomposition naturally defines a graph structure on the space of Hilbert-space factors, where each node corresponds to a factor and two nodes are connected by an edge if they have a nonzero interaction in the Hamiltonian.
To go from this topological structure to a geometric one, we need to look beyond the Hamiltonian to the specifics of an individual low-lying state.
Given any factor of Hilbert space constructed from a collection of smaller factors, $\HH_A = \otimes_{\alpha \in A} \HH_\alpha \subset \HH_\R$, and its relative complement $\HH_{\bar A} = \HH_\R \backslash \HH_A$, we can construct its density matrix and entropy,
\be
  \rho_A = \Tr_{\bar A} \rho_\R\ ,\qquad S_A = - \Tr \rho_A \log \rho_A,
\ee
and given any two such factors $\HH_A$ and $\HH_B$ we can define their mutual information
\be
  I(A\co B) = S_A + S_B - S_{AB}.
\ee
Guided again by what we know about quantum field theory, we consider ``redundancy-constrained'' states, which capture the notion that nearby degrees of freedom are highly entangled, while faraway ones are unentangled.
In that case the entropy of $\rho_A$ can be written as the sum of mutual informations between micro-factors inside and outside $\HH_A$,
\be
  S_A = \frac{1}{2} \sum_{\alpha\in A, \beta \in \bar A} I(\alpha \co \beta).
\ee

The mutual information allows us to assign weights to the various edges in our Hilbert-space-factor graph.
With an appropriate choice of weighting, these weights can be interpreted as distances, with large mutual information corresponding to short distances \cite{Cao:2016mst}.
That gives our graph an emergent spatial geometry, from which we can find a best-fit smooth manifold using multidimensional scaling.
(Alternatively, the entropy across a surface can be associated with the surface's area, and the emergent geometry defined using a Radon transform \cite{Cao:2017hrv}.)
As the quantum state evolves with time according to the Schr\"odinger equation, the spatial geometry does as well; interpreting these surfaces as spacelike slices with zero extrinsic curvature yields an entire spacetime with a well-defined geometry.

\begin{itemize}
\item \textbf{Emergent classicality}
\end{itemize}

A factorization of Hilbert space into local micro-factors is not quite the entire story.
To make contact with the classical world as part of an emergent description, we need to further factorize the degrees of freedom within some region into macroscopic ``systems'' and a surrounding ``environment," and define a preferred basis of ``pointer states'' for each system. 
This procedure is crucial to the Everettian program, where the interaction of systems with their environment leads to decoherence and branching of the wave function.
To describe quantum measurement, one typically considers a quantum object $\HH_q$, an apparatus $\HH_a$, and an environment $\HH_e$.
Branching occurs when an initially unentangled state evolves first to entangle the object with the apparatus (measurement), and then the apparatus with orthogonal environment states (decoherence), for example:
\begin{align}
  |\psi\rangle &= (\alpha|+\rangle_q + \beta|-\rangle_q)\otimes |0\rangle_a \otimes |0\rangle_e \\
  & \rightarrow (\alpha|+\rangle_q|+\rangle_a + \beta|-\rangle_q|-\rangle_a) \otimes |0\rangle_e \\
  & \rightarrow \alpha|+\rangle_q|+\rangle_a|+\rangle_e + \beta|-\rangle_q|-\rangle_a|-\rangle_e .
\end{align}
The Born Rule for probabilities, $p(i) = |\psi_i|^2$, isn't assumed as part of the theory; it can be derived using techniques such as decision theory\cite{2009arXiv0906.2718W} or self-locating uncertainty \cite{Sebens:2014iwa}.

Two things do get assumed: an initially unentangled state, and a particular factorization into object/apparatus/environment.
The former condition is ultimately cosmological -- the universe started in a low-entropy state, which we won't discuss here.
The factorization, on the other hand, should be based on local dynamics.
While this factorization is usually done based on our quasi-classical intuition, there exists an infinite unitary freedom in the choice of our system and environment. 
We seek an algorithm for choosing this factorization that leads to approximately classical behavior on individual branches of the wave function.

This question remains murky at the present time, but substantial progress is being made.
The essential observation is that, if quantum behavior is distinguished from classical behavior by the presence of entanglement, classical behavior may be said to arise when entanglement is relatively unimportant.
In the case of pointer states, this criterion is operationalized by the idea that such states are the ones that remain robust under being monitored by the environment \cite{Zurek:1981xq}.
For a planet orbiting the Sun in the solar system, for example, such states are highly localized around classical trajectories with definite positions and momenta.

A similar criterion may be used to define the system/environment split in the first place \cite{Tegmark:2014kka,mereology}.
Consider a fixed Hamiltonian and some Hilbert-space factorization into subsystems $A$ and $B$. 
Generically, if we start with an unentangled (tensor-product) state in that factorization, the amount of entanglement will grow very rapidly.
However, we can seek the factorization in which there exist low-entropy states for which entanglement grows at a minimum rate.
That will be the factorization in which it is useful to define robust pointer states in one of the subsystems, while treating the other as the environment.

This kind of procedure for factorizing Hilbert space is, in large measure, the origin of our notion of preferred classical variables.
Given a quantum system in a finite-dimensional part of Hilbert space, in principle we are able to treat any Hermitian operator as representing an observable.
But given the overall Hamiltonian, there will be certain specific interaction terms that define what is being measured when some other system interacts with our original system.
We think of quantum systems as representing objects with positions and momenta because those are the operators that are most readily measured by real devices, given the actual Hamiltonian of the universe.
We think of ourselves as living in position space, rather than in momentum space, because those are the variables in terms of which the Hamiltonian appears local.

\begin{itemize}
\item \textbf{Gravitation from entanglement}
\end{itemize}

We have argued that the geometry of spacetime can be thought of as arising from the entanglement structure of the quantum state in an appropriate factorization.
To match our empirical experience of the world, this emergent spacetime should respond to emergent energy-momentum through Einstein's equation of general relativity.
While we can't do full justice to this problem in this essay, we can mention that there are indications that such behavior is quite natural.

The basic insight is Jacobson's notion of ``entanglement equilibrium" \cite{Jacobson:2015hqa}, extended to the case where spacetime itself is emergent rather than postulated \cite{Cao:2017hrv}.
Consider a subsystem in Hilbert space, in a situation where the overall quantum state is in the vacuum.
It is then reasonable to imagine that the subsystem is in entanglement equilibrium: a small perturbation leaves the entropy of the region unchanged to first order.
If we divide the entanglement into a small-scale ultraviolet term that determines the spacetime geometry, and a longer-scale infrared term characterizing matter fields propagating within that geometry, the change in one kind of entropy must be compensated for by a corresponding change in the other,
\be
  \delta S_{UV} = - \delta S_{IR}.
\ee
Here the left-hand side represents a change in geometry, and can be related directly to the spacetime curvature.
The right-hand side represents a matter perturbation, which can be related to the modular Hamiltonian of an emergent effective field theory on the background.
At the linearized level (the weak-field limit), it can be shown that this relation turns into the $00$ component of Einstein's equation in the synchronous gauge,
\be
  \delta G_{00} = 8\pi G \delta T_{00}.
\ee
If the overall dynamics are approximately Lorentz invariant (which they must be for this program to work, although it's unclear how to achieve this at this time), demanding that this equation hold in any frame implies the full linearized Einstein's equation, $\delta G_{\mu\nu} = 8\pi G \delta T_{\mu\nu}$.

There are a number of assumptions at work here, but it seems plausible that the spacetime dynamics familiar from general relativity can arise in an emergent spacetime purely from generic features of the entanglement structure of the quantum state.
Following our quantum-first philosophy, this would be an example of finding gravity within quantum mechanics, rather than quantizing a classical model for gravitation.

\newpage

\begin{itemize}
\item \textbf{The problem(s) of time}
\end{itemize}

Given our ambition to find the most minimal fundamental description of reality, it is natural to ask whether time as well as space could be emergent from the wave function.
The Wheeler-deWitt equation of canonical quantum gravity takes the form
\be
  \hat{H}|\psi\rangle = 0,
\ee
for some particular form of $\hat H$ in a particular set of variables.
In this case time dependence is absent, but one may hope to recover an emergent notion of time by factorizing Hilbert space into a ``clock'' subsystem and the rest of the universe, 
\be
  \HH = \HH_U \otimes \HH_C,
  \label{clock}
\ee
then constructing an effective Hamiltonian describing evolution of the universe with respect to the clock.

Given our discussion thus far, the problem with such a procedure should be clear: what determines the decomposition (\ref{clock})?
In the Schr\"odinger case we can have data in the form of the spectrum of the Hamiltonian, but in the Wheeler-deWitt case the universe is in a single eigenstate; no other features of the Hamiltonian, including its other energy eigenvalues, can be relevant.
This problem has been dubbed the ``clock ambiguity" \cite{Albrecht:2007mm}.

One potential escape would be to imagine that the fundamental state of the universe is described not by a vector in Hilbert space, but by a density operator acting on it.
Then we have an alternative set of data to appeal to: the eigenvalues of that density matrix.
These can be used to compute a modular Hamiltonian (given by the negative of the logarithm of the density operator), which in turn can yield an effective notion of time evolution, a proposal known as the ``thermal time hypothesis" \cite{Connes:1994hv}.
Thus it is conceivable that time as well as space could be emergent, at the cost of positing a fundamental density operator describing the state of the universe.\footnote{If time is fundamental rather than emergent, there is a very good reason to believe that the entirety of Hilbert space is infinite-dimensional, even if the factor describing our local region is finite-dimensional; otherwise the dynamics would be subject to recurrences and Boltzmann-brain fluctuations \cite{Carroll:2008yd}.}

\begin{itemize}
\item \textbf{Prospects and puzzles}
\end{itemize}

The program outlined here is both ambitious and highly speculative.
We find it attractive as a way of deriving most of the familiar structure of the world from a minimal set of truly quantum ingredients.
In particular, we derive rather than postulating such notions as space, fields, and particles.
The fact that our Hilbert space is locally finite-dimensional suggests an escape from the famous problems of ultraviolet divergences in quantum field theory, and the emergence of spacetime geometry from quantum entanglement is an interesting angle on the perennial problems of quantum gravity.

Numerous questions remain; we will highlight just two.
One is the emergence of local Lorentz-invariant dynamics.
There are no unitary representations of the Lorentz group on a finite-dimensional factor of Hilbert space.
This might seem to imply that Lorentz symmetry would be at best approximate, a possibility that is experimentally intriguing but already highly constrained.
It would be interesting to understand how numerically large any deviations from perfect Lorentz invariance would have to be in this framework, and indeed if they have to exist at all.

The other issue is the emergence of an effective field theory in curved spacetime that could describe matter fields in our geometric background.
We have posited that a field theory might be identified with infrared degrees of freedom while the geometry is determined by ultraviolet degrees of freedom, but there is much to be done to make this suggestion more concrete.
A promising idea is to invoke the idea of a quantum error-correcting code \cite{Harlow:2016vwg,Cao:2017hrv}.
Such codes imagine identifying a ``code subspace'' within the larger physical Hilbert space, such that the quantum information in the code can be redundantly stored in the physical Hilbert space.
There is a natural way to associate the code subspace with the infrared degrees of freedom of the matter fields, with the rest of the physical Hilbert space providing the ultraviolet entanglement that defines the emergent geometry.
Once again, this is a highly speculative but a promising line of investigation.

We are optimistic that this minimal approach to the ontology of quantum mechanics is sufficient, given an appropriate Hamiltonian and quantum state, to recover all of the richness of the world as we know it.
It would be a profound realization to ultimately conclude that what is fundamental does not directly involve spacetime or propagating quantum fields, but simply a vector moving smoothly through a very large-dimensional Hilbert space.
Further investigation will be needed to determine whether such optimism is warranted, or whether we have just gone mad.

\begin{center} 
 {\bf Acknowledgments}
 \end{center}

We are thankful to ChunJun (Charles) Cao for helpful conversations.
This research is funded in part by the Walter Burke Institute for Theoretical Physics at Caltech and by DOE grant DE-SC0011632.

\bibliographystyle{utphys}
\bibliography{qmessayrefs}

\end{document}